\documentclass[12pt,showpacs,amsmath,amssymb]{revtex4}%
\usepackage{color}
\usepackage{epic}
\usepackage{eepic}
\usepackage{graphicx}
\usepackage{bm}

\usepackage{graphicx}
\usepackage{dcolumn}
\usepackage{bm}
\usepackage[hyperfootnotes=false]{hyperref}
\hypersetup{
  colorlinks   = true, 
  urlcolor     = blue, 
  linkcolor    = blue, 
  citecolor   = red 
}

\begin{document}

\title{The Model of Quantum Thermodynamics From the First Principles:
Quantum Halo or Small Environment }

\author{Ashot Gevorkyan}
\altaffiliation{$^1$Institute for Informatics and Automation Problems NAS of RA,\\
$^2$Institute of Chemical Physics  NAS of RA }\quad
\email{g\_ashot@sci.am}

\date{\today}
\begin{abstract}
The evolution of the joint system (JS) - ``\emph{quantum system} (QS)+\emph{thermal bath}
(TB)" is considered in the framework of a complex probabilistic processes that satisfies
the stochastic differential equation of the Langevin-Schr\"{o}dinger type. Two
linearly coupled oscillators that randomly interact with the environment and
with each other are selected as QS. In the case when the interactions obey
the law of a white random process, all the construction of the statistical
parameters of the QS and its environment are performed analytically in the
form of double integrals and solutions of second-order partial differential
equations. Expressions of time-dependent von Neumann entropy and its generalization
are obtained, taking into account the self-organization and entanglement processes
occurring in the JS. It is mathematically proved that as a result of the relaxation
of JS in the TB, a \emph{small quantized environment} is formed, which can be interpreted
as a continuation of QS or its \emph{halo}. Bell states formed as a result of
the decay of coupled two linear oscillators are constructed taking into account
the influence of the environment. The transitions between $(in)$ and $(out$)
asymptotic states of QS are studied in detail taking into account the influence
of TB. Within the framework of the model problem, the possibility of constructing
quantum thermodynamics from the first principle is proved without using any additional
conditions.
\end{abstract}

\maketitle
\section{I\lowercase{ntroduction}}
\label{01}
When we try to strictly and consistently approach the study of a \emph{quantum system} (QS),
it becomes obvious that its isolation from the environment is an almost nonrealizable task.
Note that even if we assume that it is possible to exclude the interaction of QS with the
environment, taking into account the influence of quantum fluctuations of vacuum in any
case makes the QS an open system \cite{Dav,Dav1,PWM,Gev,ASG,ASG1}. In other words, a full
description of QS also requires the inclusion of its environment, which is essentially
the whole universe. Thus, the fact that any QS in a certain sense is an open system means
that in reality not a single quantum state can be a pure state.

In recent years investigation of problems of influence of an environment on properties of
the quantum system has become a subject of great interest in view of importance of solutions
a set of fundamental and applied problems of science and technologies, including the quantum
foundations of thermodynamical behavior \cite{Gem,Bor}, the nature of the quantum measurement
process \cite{Allahverdyan}, dissipation and decoherence \cite{Devis,Breuer,Marq,Zurek},
the question of how the classical properties emerges from the quantum world \cite{Sch},
the problems of quantum correlations and quantum discord \cite{Zur}, the peculiarities of the
entanglement at the Bose condensation \cite{Joshi,Dun}  etc.

Usually, when we speak about the environment, as a rule, we mean a large system (\emph{thermostat-
thermal bath} (TB)), which in a state of thermodynamical equilibrium is characterized by a certain
temperature and distributions of various physical parameters. Recall that the theory of open quantum
systems (OQS) \cite{Br,Lind}, which is the basis of almost all modern studies in the field of quantum mechanics
and its applications, takes into account the influence of the thermostat on QS, while the effect
of QS on the thermostat is not taken into account. This assumption, which underlies the theory
of OQS, in many cases is justified, nevertheless, there are cases when the effects of the openness
of quantum systems are not only impossible to neglect, but they must be taken into account with
all mathematical rigor. Note that this is especially relevant in the field of quantum information
processing, the advantage of which is obvious in comparison with classical information processing \cite{Niel}.
Mathematically, a more rigorous and consistent formulation of OQS is very important for studying
the relaxation of not only QS, but also the thermostat itself in order to avoid the loss of information
 inherent in open systems.

Recently, when studying the properties of quantum thermodynamics in the framework of model
calculations, it was shown that thermal behavior manifests itself even with sufficiently small
entangled quantum systems  \cite{Gem,Bor}. In particular, calculations show that micro-canonical
and thermal behavior have been observed, including the rapid approach to micro-canonical equilibrium
entropy, that is typical to canonical systems. The results of these numerical simulations are explained
theoretically \cite{Popescu}, wherein shown the key role of the quantum entanglement between the
system and its environment in  attainment of thermodynamic equilibrium in the system. Moreover, it
is proved that the basic postulate of statistical mechanics, namely \emph{the postulate of
a priori equal  probability of statistical states}, should be discarded as unnecessary and misleading.
In \cite{Kellman}, a model system of coupled quantum oscillators (CQO) was calculated by way of
numerical simulation, which interacts with a  quantum environment.

The purpose of this article is to develop an analytical approach that will allow us to study
the evolution of a \emph{quantum system and its environment}  as a \emph{closed inextricable system }
depending on time. The main idea on which the developed approach is based is as follows:
we assume that during the evolution of the quantum subsystem, i.e. CQO, it is subjected
to continuous random environmental influences.  In this case, the quantum state of the
joint system (CQO+thermostat) can be well described  using \emph{complex probabilistic process}
(CPP) that satisfies a stochastic differential equation of the Langevin-Schr\"{o}dinger type.

Note that in the work, using an equation of the Langevin-Schr\"{o}dinger type as the first
principle, we construct  all the parameters and the corresponding distributions characterizing
the quantum subsystem and its so-called a \emph{small environment}.

The manuscript is organized as follows:

In Section II, the problem of coupled quantum oscillators is formulated taking into account the
presence of a random environment. The explicit form of the wave function of the \emph{joint system}
(JS)  `` CQO + thermostat '' is obtained by solving the  Langevin-Schr\"{o}dinger
equation in the form of an orthonormal probability processes.

In Section III, a method of \emph{stochastic density matrix} (SDM) has been developed.
The section contains the basic definitions with which the statistical parameters of the
quantum subsystem and small environment are constructed.

In Section IV, we obtain stochastic equations describing a thermostat under various conditions
and equations for distributions of the corresponding fields of an environment under conditions
of thermodynamic equilibrium.

Section V is devoted to calculating quantum entropy. In particular, the von Neumann entropy
for the “ground state” is analyzed in detail and its generalization is given.

In Section VI, the energy levels and their populations are constructed in the form of two-dimensional
 integral representations.

Section VII considers the problem of entanglement of two oscillators as a result of relaxation of
CQO in the thermostat. The formation of Bell states is considered in detail.An equation is also derived
that describes the evolution of the thermostat under the influence of CQO and the formation of
 the so-called \emph{small environment} (SE).

In Section VIII, a general expression is constructed for the transition probabilities between
different  $(in) $ and $(out)$ asymptotic state, and an exact method for its calculation is developed.

In Section IX, the obtained results are discussed in detail and further ways of development of
the problems under consideration are indicated.

\section{F\lowercase{ormulation of the problem}}
\label{02}
We consider the quantum subsystem and the random environment as a \emph{joint system}
(JS), which is described in the framework of the \emph{stochastic differential equation}
(SDE) of the Langevin-Schr\"{o}dinger type (L-Sch):
\begin{equation}
i\partial_{t}\Psi_{stc} =\mathrm{{\hat H}}\bigl({\bf x},t;\{{\bf f}\}\bigr) \,
\Psi_{stc},\qquad  \partial_t\equiv\partial/\partial t,
\label{1.01}
\end{equation}
where  ${\bf x}=(x_1,x_2)\in (-\infty,+\infty)$ and $t\in(-\infty,+\infty)$.

For definiteness, we will study the evolution of a subsystem consisting of two linearly
coupled 1$D$ quantum harmonic oscillators immersed in a random environment (thermostat).
The evolution operator $\mathrm{{\hat H}}({\bf x},t;\{{\bf f}\})$  in the units
$\hbar=m=1$ we will present in the form:
\begin{eqnarray}
\mathrm{{\hat H}}\bigl({\bf x},t;\{{\bf f}\}\bigr) = \frac {1}{2}
\sum_{l=1}^2\Bigl[-\frac {\partial^2}{\partial
x^2_l}+ \Omega^2\bigl(t;\{{\bf f}\}\bigr)x^2_l\Bigr]+
\mathcal{\omega}\bigl(t;\{{\bf f}\}\bigr)x_1x_2,
 \label{1.02}
\end{eqnarray}
where it is assumed that  $\Omega\bigl(t;\{{\bf f}\}\bigr)$ and
$\omega\bigl(t;\{{\bf f}\}\bigl)$ are arbitrary time functions consisting of regular
and random components.

We define JS randomness as a complex probabilistic process $\{{\bf f}\}$,
which will be clearly defined below (see (\ref{1.07})). For further analytical
study of the problem, it is important to bring the operator  (\ref{1.02}) into
a diagonal form.

By performing the following coordinate transformations:
\begin{equation}
q_1=\frac{x_1-x_2}{\sqrt{2}}, \qquad q_2=\frac{x_1+x_2}{\sqrt{2}},
\label{1.03}
\end{equation}
in the equation (\ref{1.02}) we get:
\begin{equation}
{\hat{\mathcal{H}}}\bigl(q_1,q_2,t;\{{\bf f}\}\bigr) = \frac
{1}{2}\sum_{l=1}^2\Bigl[-\frac {\partial^2}{\partial q^2_l}+
\Omega^2_l\bigl(t;\{{\bf f} \}\bigr)q^2_l\Bigr],
\label{1.04}
\end{equation}
where $\Omega_l\bigl(t;\{{\bf f} \}\bigr)$ denotes the effective frequency:
\begin{equation}
\Omega^2_l\bigl(t;\{{\bf f}\}\bigr)=\Omega^2(t;\{{\bf f}\}\bigr)-(-1)^l
\omega\bigl(t;\{{\bf f}\}\bigr)\geq 0.
 \label{1.05}
\end{equation}
In particular, as follows from (\ref{1.05}), the restriction
$|\omega\bigl(t;\{{\bf f}(t)\}\bigr)|\leq Re[\Omega^2\bigl(t;\{{\bf f}(t)\}\bigr)]$,
is imposed on the function $\omega\bigl(t;\{{\bf f}(t)\}\bigr)$.
For definiteness,  we will represent the effective frequency as a sum:
\begin{equation}
\Omega^2_l\bigl(t;\{{\bf f}\}\bigr)=\Omega_{0l}^2(t)+f_l(t),\qquad
\{{\bf f}(t)\}=\bigl(f_1(t),f_2(t)\bigl).
 \label{1.05a}
\end{equation}
Note that $\Omega_{0l}(t)$ is a real function, and $f_l(t)=f_l^{(r)}(t)+ if_l^{(i)}(t)$
is a random complex function, where $f_l^{(r)}$ and $f_l^{(i)}(t)$ denote its real and
imaginary parts, respectively. As for the function $f_l(t)$, it describes the influence
of the environment or TB on the QS, which consists of real
$f^{(r)}_l$ and imaginary $f^{(i)}_l$ parts. Note that in a quality of an environment
model is often used the set of harmonic oscillators \cite{FV,Dekk,Cald,Joos,Paz}
that is equivalent to the quantized fields \cite{Unhru,Zurek1}.

Note that in the absence of a random environment, we turn to the usual problem of a
parametric quantum oscillator with a regular Hamiltonian:
\begin{equation}
 {\hat{\mathcal{H}}}_0\bigl(q_1,q_2,t\bigr)={\hat{\mathcal{H}}}\bigl(q_1,q_2,t;\{{\bf f}\}\bigr)
 \Bigl|_{\{{\bf f}\}\equiv\,0}\,= \frac{1}{2}\sum_{l=1}^2\biggl[-\frac{\partial^2}{\partial q^2_l}
 +\Omega^2_{0l}(t)q^2_l\biggr].
 \label{1.b05a}
\end{equation}

We will suppose that the  functions $\Omega_{0l}(t)$ and $f_l(t)$ satisfy the following
asymptotic conditions:
\begin{eqnarray}
\lim_{t\to\,\mp\,\infty}\Omega_{0l}(t)=\Omega_{l}^{\mp}=const_\mp>0,\qquad
\lim_{t\to\,-\,\infty}f_l(t)=0. \label{1.06}
\end{eqnarray}
For further analytical constructions, we will assume that $f^{(\upsilon)}_l(t)$, where
$\upsilon=(r,i)$, is an independent Gaussian  process with zero mean value and a delta-like
correlation function:
\begin{equation}
 \langle f^{(\upsilon)}_l(t)\rangle=0,\qquad
\langle f^{(\upsilon)}_l(t)f^{(\upsilon)}_l(t')\rangle = 2\epsilon^{(\upsilon)}_l\,\delta (t-t').
\label{1.07}
\end{equation}
Note that $\epsilon^{(\upsilon)}_l$ denotes the power of environment fluctuations (TB fluctuations),
which is natural for gases to assume that $\epsilon^{(r)}_l=\bar{\epsilon}_l=kT_l$. As regards to
the constant $\epsilon^{(i)}_l$, then it characterizes a dissipation processes  of  QS in the TB and
it is assumed that $\epsilon^{(i)}_l=\mu\epsilon^{(r)}_l $, where $0\leq\mu\leq 1$ is a some constant,
in addition, $k$ and $T_l$ are the Boltzmann constant and temperature TB, respectively. In the case
when $\epsilon_1\neq\epsilon_2$ and $\epsilon_l=(\epsilon^{(r)}_l,\epsilon_l^{(i)})$, the TB
in thermodynamic equilibrium is characterized by two temperatures.

It is easy to show that the equation (\ref{1.01}) in the limit $t\rightarrow-\infty$
 or in the $(in)$ asymptotic state has the following factorized solution:
\begin{equation}
\Phi_{in}(\textbf{n}|\textbf{q},t)
=\prod_{l=1}^{2}e^{-i(n_l+1/2)\,\Omega_{l}^{-}t}\phi_{n_l}(q_l), \qquad n_l=0,1,...,
\label{1.08}
\end{equation}
where $\textbf{n}\equiv(n_1,n_2)$ and $\textbf{q}\equiv(q_1,q_2)$, in addition:
\begin{equation}
\phi_{n_l}(q_l)= \biggl(\frac{g_l^-}{2^{n_l}{n_l}!}\biggr)^{1/2}
e^{-({\Omega_{l}^{-}q_l^2})/{2}}\mathrm{H}_{n_l} \Bigl(\sqrt{\Omega_{l}^{-}}q_l\Bigr),
\qquad g_l^-=({\Omega_{l}^{-}}/{\pi})^{1/2}.
 \label{1.09}
\end{equation}
Recall that $\phi_{n_l}(q_l)$ denotes the wave function of $1D$ stationary quantum
harmonic oscillator, and $\mathrm{H}_{n_l}(q_l)$ is the Hermitian polynomial.

Let us represent a general solution of the problem (\ref{1.01})-(\ref{1.09}) in factorized form:
 \begin{eqnarray}
\Psi_{stc}(\textbf{q},t)=\prod_{l=1}^{2}Y_{stc}^{(l)}(q_l,t),
 \label{1.10}
\end{eqnarray}
where $Y_{stc}^{(l)}(q_l,t)$  denotes the stochastic wave function, the solution of $1D$
quantum harmonic oscillator for an arbitrary time-dependent frequency  $\Omega_l(t)$.

The explicit form of this wave function is well known (see for example \cite{Baz}):
\begin{eqnarray}
Y^{(l)}_{stc}(q_l,t) =\frac {1}{\sqrt{\sigma_l}}\exp\biggl
\{\frac{i}{2}\frac{\dot{\sigma}_l}{\sigma_l}q^2_l\biggr\}\chi_l,\qquad
\dot{\sigma}_l(t)=\frac{d\sigma_l(t)}{dt},
 \label{1.11}
\end{eqnarray}
where $\chi_l$  is the solution of 1$D$ Schr\"{o}dinger equation for the autonomous
 harmonic oscillator on a stochastic space-time continuum $\{y_l,\tau_l\}$:
\begin{equation}
\label{1.12} i\frac {\partial \chi_l }{\partial \tau_l}=\frac{1}
{2}\Bigl[-\frac {\partial^2}{\partial
y_l^2}+{(\Omega^{-}_{l}y_l)}^2\Bigr]\,\chi_l.
\end{equation}
Note that the following notations are made in (\ref{1.12}):
$$
y_l=\frac{q_l}{\sigma_l},\,\,\quad\xi_l(t)=\sigma_l(t)e^{i\gamma_l(t)},\,\,\quad
\tau_l =\frac{\gamma_l (t)}{\Omega_{l}^-},\,\,\quad \gamma_l(t)=\Omega_
l^-\int_{-\infty}^{\,t}\frac{dt'}{\sigma^2_l(t')},
$$
where the function $\xi_l(t)$  is a solution to the classical oscillatory equation:
\begin{equation}
\ddot \xi_l +\Omega^2_l(t)\xi_l=0,\qquad \dot \xi_l=d\xi_l/dt.
\label{1.13}
\end{equation}

Now, taking into account the expressions (\ref{1.10})-(\ref{1.12}), we can construct
the wave function of the JS:
$$
\Psi_{stc}(\textbf{n}|\textbf{q},t,\{\bm\xi\})= \prod_{l=1}^2Y^{(l)}_{stc}(n_l|q_l,t;\{\xi_l\}),
$$
where $\{\bm\xi\}\equiv\{\xi_1(t),\xi_2(t)\}$ and $\textbf{n}=(n_1,n_2)$, in addition:
\begin{eqnarray}
Y^{(l)}_{stc}=\biggl(\frac {g_l^-}{2^{n_l}n_l!\,\sigma_l}\biggr)^{1/2}
\exp\biggl\{\frac{1}{2}\Bigl(i\frac{\dot{\sigma_l}}{\sigma_l}-\frac{\Omega_{l}^-}
{\sigma_l^2}\Bigr)\,q_l^2-i\,\Bigl(n_l+\frac12\Bigr)\,\Omega_{l}^-\int_{-\infty}^{\,t}
\frac{dt'}{\sigma_l^2}
 \biggr\}\mathrm{H}_{n_l}\Bigl(\sqrt{\Omega_{l}^-}\frac{q_l}{\sigma_l}\Bigr).
\nonumber\\
\label{1.14}
\end{eqnarray}
It is easy to show that the  wave functions of JS in the Hilbert space form a full orthonormal basis:
\begin{equation}
\int\Psi_{stc}(\textbf{n}|\textbf{q},t;\{\bm\xi\})
\Psi_{stc}^{\ast}(\textbf{m}|\textbf{q},t;\{\bm\xi\})d\textbf{q}=\prod_{l=1}^{2}\delta_{n_lm_l},
 \label{1.15}
\end{equation}
where   $\delta_{n_lm_l}$  is the Kronecker symbol.

Note that the wave function $\Psi_{stc}(\textbf{n}|\textbf{q},t;\{\bm\xi\})\in
L_2(\Xi)$, in addition, $\Xi\cong\mathbb{R}^2\otimes\mathbb{R}_{\{\bm\xi\}}$ is the
extended  space, where $\mathbb{R}^2$  is  2$D$ Euclidean space, while $\mathbb{R}_{\{\bm\xi\}}$
denotes a functional or uncountable-dimensional space, the measure of which is exactly
determined (see \cite{}).


\section{T\lowercase{he free evolution of an environment fields}}
\label{sec-3}
The solution of the classical equation (\ref{1.13})
can  be represented in the form:
\begin{equation}
\xi_l(t)= \Biggl\{
\begin{array}{ll}
\qquad\xi_{0l}(t),\qquad\qquad\qquad t\leq t_0,
\vspace{0.2 cm}\\
\xi_{0l}(t_0)\exp{\bigl\{\Theta_{l}(t)\bigr\}},\,\,\qquad t>t_0,
\end{array}
\label{3.01}
\end{equation}
where $\xi_{0l}(t)$ is the solution of the classical  equation (\ref{1.13})
for the regular frequency $\Omega_{0l}(t)$ and $\Theta_{l}(t)=\int^t_{t_0}\phi_{l}(t')\,dt'$,
in addition, $t_0$ denotes the time of switching on of a random environment or TB.

Substituting (\ref{3.01}) into (\ref{1.13}), we get the following
nonlinear complex SDE:
\begin{equation}
\label{3.02} \dot{\phi_{l}}+\phi_{l}^2+{\Omega^2_{0l}}(t)+f_l(t)=0,
\end{equation}
where $\dot{\phi_{l}}=d\phi_{l}/dt$ and $\phi_{l}(t_0)=i\Omega^-_{l}$.

For further study, the complex function $\phi_l(t)$ is conveniently represented as a sum:
\begin{equation}
\label{3.03} \phi_{l}(t)=u^{(l)}_{1}(t)+iu^{(l)}_{2}(t).
\end{equation}
Using (\ref{3.03}) and (\ref{3.02}), we can write the following
system of nonlinear SDEs \cite{Gev}:
\begin{equation}
 \Biggl\{
\begin{array}{ll}
 \dot{u}^{(l)}_{1}=\bigl(u^{(l)}_{2}\bigr)^2-\bigl(u^{(l)}_{1}\bigr)^2-
 \Omega_{0l}^2(t)-f^{(r)}_l(t),\,\\
\dot u^{(l)}_{2}=-2u^{(l)}_{1}u^{(l)}_{2}-f_l^{(i)}(t), \qquad\qquad\qquad\qquad
\end{array}
\label{3.04}
\end{equation}
where the fields
${\bf u}(t)=\bigl\{[u_1^{(1)}(t), u_2^{(1)}(t)];\,[u_1^{(2)}(t),u_2^{(2)}(t)]\bigr\}$
satisfy the following asymptotic conditions:
$$
\dot{u}^{(l)}_{1}(t_0)={Re[\dot \xi_l(t_0)/\xi_l(t_0)]}=0,\qquad
 \dot{u}^{(l)}_{2}(t_0)=Im[\dot
\xi_l(t_0)/\xi_l(t_0)]=\Omega_{l}^-.
$$
Note that depending on the character of random forces $f_1(t)$ and $f_2(t)$ the
probabilistic processes (fields) $\phi_{1}(t)$ and $\phi_{2}(t)$ can be implemented
by way of two different scenarios.

First, we consider a scenario when random forces are independent. In this case,
it is obvious that the distribution of the fields of TB can be represented in
factorized form:
\begin{equation}
\mathcal{P}(\textbf{u},t\vert \textbf{u}_0,t_0 ) = \prod_{l=1}^2\Bigl
<\prod_{j=1}^2\delta \bigl(u_j^{(l)}(t)-u^{(l)}_{0j}\bigr)\Bigr>,
 \label{3.05a}
\end{equation}
where $u^{(l)}_{0j}=u_j^{(l)}(t_0)$ is a component of fields in the state
$(in)$. To simplify the notation in the equations below, the superscripts
of the random fields will be omitted.

Using SDE (\ref{3.04}) for the probability distribution of fields, we can
find the following Fokker-Planck type equation (see in detail \cite{Gred,Gard}):
\begin{equation}
\partial_t P_{l}=\hat L_{l}\bigl(\Omega_{0l}(t),\epsilon_l|u_1,u_2,t\bigr)P_{l},
\label{3.05}
\end{equation}
where the evolution operator has the following form:
$$
\hat L_{l}=\bar{\epsilon}_l \biggl(\frac {\partial^{\,2}}{\partial u_1^2}\,
+\mu\frac{\partial^{\,2}}{\partial u_2^2}\biggr)\,+ \frac {\partial
}{\partial u_1}\bigl(u_1^2-u_2^2+\Omega_{0l}^2(t)\bigr)+2u_1\frac
{\partial}{\partial u_2}u_2.
$$
It is quite natural to assume that the solution of the equation (\ref{3.05})
satisfies the initial conditions:
\begin{eqnarray}
\label{3.06}
P_{l}(u_1,u_2,t_0)=\prod_{j=1}^2\delta(u_j-u_{0j}),
 \qquad
P_{l}(u_1,u_2,t)\bigl|_{||\textbf{u}||\rightarrow\,\infty}\,\,\rightarrow
0,\qquad\,\,\,
\end{eqnarray}
where $||\textbf{u}||=(u_1^2+u_2^2)^{1/2}$, in addition, $ P_{l}(u_1,u_2,t)$
as a function of the probability density of the environmental fields should
be normalized to unity.

Regarding the physical meaning of solving the equation (\ref{3.05}), it is
easy to verify that this is the distribution of the fields of the so-called
\emph{small environment} (SE) formed under the  influence of the classical
oscillator with the frequency $\Omega_{0l}(t)$.

When implementing the second scenario, we assume that the source of random
forces is the same and, accordingly, the fields distribution can be defined as:
\begin{equation}
\mathcal{P}(\textbf{u},t\vert \textbf{u}_0,t_0)=
\biggl<\prod_{l=1}^2\prod_{j=1}^2\delta\bigl(u_j^{(l)}(t)-u^{(l)}_{0j}\bigr)\biggr>,
  \label{3.07}
\end{equation}
which is a solution to the following equation:
\begin{equation}
 \partial_t \mathcal{P}=\sum_{l=1}^2\hat L_{l}\bigl(\Omega_{0l}(t),\epsilon_l|u_1,u_2,t\bigr)
 \mathcal{P}.
  \label{3.08}
\end{equation}
The equation (\ref{3.08}) can be solved using initial conditions similar to (\ref{3.06}).

Recall that the equations (\ref{3.05}) and (\ref{3.08}) describe the distributions of
environmental fields (\emph{small environment}), when the first and second scenarios
are realized, respectively. Recall that in the implementation of the first scenario,
the environmental fields are generated by two independent random forces, whereas in
the case of the second scenario, these random forces have a common nature. Below we
will study the statistical properties of QS and the environment associated with the
realization of the first scenario.

\section{M\lowercase{ethod of the stochastic density matrix}}
To study irreversible processes in quantum systems, the representation of a
nonstationary density matrix developed in the framework of the Liouville-von
Neumann equation is often used \cite{Neumann}. However, there is an important
limitation associated with the  application of this representation, which in number
of cases can strongly distort the described phenomena and expected results.
Note that the representation will be consistent and rigorous if it allows one to take
into account both the influence of the environment on the QS and the impact of
the QS on the environment. As noted above, this can lead to the formation of
the so-called \emph{small environment}, which will have interesting physical
properties and have a specific geometric and topological features. Unfortunately,
both the standard representation for the density matrix and many modern approaches
describing relaxation processes occurring in open systems do not allow achieving
the specified rigor of description. To eliminate this flaw in the theory, we
propose a new approach, the \emph{stochastic density matrix} (SDM) method, which
allows us to describe a JS, taking into account the self-consistency between its parts.
In other words, we consider and study JS as a \emph{closed system} that is impossible
to implement within the framework of other approaches.

\textbf{Definition 1.} \emph{Let the stochastic density matrix} (SDM) \emph{be
defined as a bilinear form:}
\begin{eqnarray}
 \varrho_{stc}(\textbf{q},t;\{\bm\xi\}|\textbf{q}',t'; \{\bm\xi^{\prime}\})=
\sum_{\textbf{n}}w_{\textbf{n}}\varrho_{stc}^{(\textbf{n})}
(\textbf{q},t; \{\bm\xi\}|\textbf{q}',t';\{\bm\xi^{\prime}\}),
\label{2.01}
\end{eqnarray}
\emph{where $\varrho_{stc}^{(\textbf{n})}=\Psi_{stc}(\textbf{n}|\textbf{q},t;\{\bm\xi\})
\Psi_{stc}^{\ast}(\textbf{n}|\textbf{q}^{\,\prime},t^{\prime};\{\bm\xi^{\prime}\})$,
in addition, $w_{\textbf{n}}=w_{n_1,n_2}=w_{n_1}^{(1)}w_{n_2}^{(2)}$}
denotes the initial population of the levels of two noninteracting quantum harmonic
oscillators that, in the $(in)$ asymptotic state, respectively, possess energies;
$$
E_{n_1}=(n_1+ 1/2)\Omega_{1}^-
\qquad and \qquad
E_{n_2}=(n_2+ 1/2)\Omega_{2}^-.
$$
It is important to note that when integrating over the extended space
$\Xi$, the order of integration is important. In particular, if we first
integrate  SDM over the Euclidean space $\mathbb{R}^2$ and then over the
functional space $\mathbb{R}_{\{\bm\xi\}}$, we get:
 \begin{equation}
\label{2.02}
 \sum_{n_1,n_2=\,0}^\infty w_{n_1,n_2}=1,\qquad 0 \leq w_{n_1,n_2}\leq 1.
\end{equation}
Note that the expression (\ref{2.02}) is actually a condition for normalizing
population levels. It also means that the conservation laws are satisfied on
the extended space $\Xi$.

In the case when the integration is carried out in reverse order, that is, first
by the functional space, and then on the  Euclidean space, we obtain:
 \begin{equation}
\label{2.02a}
 \sum_{\textbf{n}} w_{\textbf{n}}\bar{\varrho}_{\textbf{n}}(T_1,T_2)=1,\qquad
\bar{\varrho}_{\textbf{n}}(T_1,T_2))\leq1,
\end{equation}
where $\bar{\varrho}_{\textbf{n}}(T_1,T_2)=
 Tr_{\bf{q}}\bigl\{Tr_{\{\bm\xi\}}\bigl[\varrho_{stc}^{(\textbf{n})}(\textbf{q},t;
\{\bm\xi\}|\textbf{q}',t'; \{\bm\xi^{\prime}\})\bigr]\bigr\}$
denotes the  population of the corresponding quantum levels at temperatures of environment
$T_1$ and $T_2$, in addition, $Tr_{\{\bm\xi\}}$ and $ Tr_{\bf{q}}$ denote integration
operations over functional  and Euclidean spaces, respectively (see below (\ref{2.04})
and (\ref{2.05})). \\
The latter means that the conservation laws in the space $\mathbb{R}^2$ as a whole are
violated, and only in the limit of statistical equilibrium  can we speak about the preservation
of their mean values.

Now we will define the mathematical expectation of various random
variables.

\textbf{Definition 2.} \emph{The reduced density matrix  is defined as the mean
value of the random density matrix:}
\begin{equation}
\label{2.03}
 \varrho(\textbf{q},t|\textbf{q}',t')=\mathbb{ E}\bigl[\varrho_{stc}\bigr]=
  Tr_{\{\bm\xi\}}\bigl[\varrho_{stc}(\textbf{q},t;
\{\bm\xi\}|\textbf{q}',t'; \{\bm\xi'\})\bigr],
\end{equation}
\emph{where $\mathbb{E}\bigl[...\bigr]$ denotes mean value of the random variable,
and $Tr_{\{\bm\xi\}}$ denotes the functional integration over the environmental
fields (see in detail \cite{AGev}):}
\begin{eqnarray}
\label{2.04}
 Tr_{\{\bm\xi\}} \bigl[K(\textbf{q},t;\{\bm\xi\}|\textbf{q}',t';\{\bm\xi'\})\bigr]=
\int  K(\textbf{q},t;{\{\bm\xi\}}|\textbf{q}',t;\{\bm\xi\})\,D\{\bm\xi\}.
\end{eqnarray}

\textbf{Definition 3.} \emph{The average value of the eigenvalue of the operator}
$\hat A(\textbf{q},t;\{\bm\xi\}|\textbf{q}',t'; \{\bm\xi'\})$
\emph{in the quantum state defined by numbers} $\textbf{n}=(n_1,n_2)$ \emph{writes as:}
\begin{equation}
\label{2.05}
 A_\textbf{n}= \lim\limits_{t\rightarrow +\infty }\Bigl\{\frac{1}{N_{\textbf{n}}(t)}
 Tr_\textbf{q}\bigl[Tr_{\{\bm\xi\}}\hat A\varrho_{stc}^{(\textbf{n}) }\bigr]\Bigr\},
\end{equation}
\emph{where} $N_{\textbf{n}}(t)=Tr_\textbf{q}\bigl[Tr_{\{\bm\xi\} }\bigl(\varrho
_{stc}^{(\textbf{n})}(\textbf{q},t;
\{\bm\xi\}|\textbf{q}',t'; \{\bm\xi^{\prime}\})\bigr)\bigr],$ \emph{ in addition:}
\begin{eqnarray}
\label{2.06} Tr_{\textbf{q}}
\bigl[K(\textbf{q},t;{\{\bm\xi\}}|\textbf{q}',t';\{\bm\xi'\})\bigr]=\int
K(\textbf{q},t;{\{\bm\xi\}}|\textbf{q}',t;\{\bm\xi'\})d\textbf{q} ,\quad
d\textbf{q}=dq_1dq_2.
\end{eqnarray}
As is known, entropy characterizes the measure of randomness of a statistical
ensemble, with the help of which one can find all the thermodynamic potentials
 of an ensemble. Recall that the entropy for the quantum system  for the first
 time was determined by von Neumann  \cite{Neumann}.

\textbf{Definition 4.} \emph{The von Neumann entropy of a quantum system is determined as:}
\begin{equation}
 \Lambda_N(t) =-Tr_{\textbf{q}}\bigl\{
\varrho(\textbf{q},\textbf{q}',t)\ln{\varrho(\textbf{q},\textbf{q}',t)}\bigr\},
 \label{2.07}
 \end{equation}
\emph{ where} $\varrho(\textbf{q},\textbf{q}',t)\equiv\varrho(\textbf{q},t|\textbf{q}',t')|_{t=t'}$
\emph{denotes the  reduced  density matrix} (RDM).

\textbf{Definition 5.} \emph{The entropy of a quantum subsystem interacting with a random
environment can be defined as follows:}
\begin{equation}
 \Lambda_G(t) =-Tr_\textbf{q}\bigl\{
 Tr_{\{\bm\xi\}}[\varrho_{stc}(\textbf{q},t;
\{\bm\xi\}|\textbf{q}',t'; \{\bm\xi^{\prime}\})\ln{\varrho_{stc}}(\textbf{q},t;
\{\bm\xi\}|\textbf{q}',t'; \{\bm\xi^{\prime}\})]\bigr\}.
 \label{2.08}
 \end{equation}
In the following, we will call this method the \emph{random entropy method} (REM).
It is assumed that the determination (\ref{2.08}) is more logical and rigorous and
suitable for cases where the quantum subsystem interacts with the environment strongly.

\section{T\lowercase{he entropy of ground state}}
Below we consider the features of relaxation immersed in a random environment
of QS. Recall that for this purpose, the study of the behavior of the entropy QS
may be the most important and informative. In this section, we will calculate the
entropy of QS in two different ways.
\subsection{The von Neumann entropy}
For simplicity, we assume that $w_{0,0}=1$ and,  accordingly, $w_{n,m}\equiv 0$
for arbitrary quantum numbers $n,m\geq1$. In this case the SDM (\ref{2.01}) can
be written in a factorized form:
\begin{eqnarray}
\label{4.01}
\varrho^{(0,0)}_{stc}(\textbf{q},t;\{\bm\xi\}|\textbf{q}',t';\{\bm\xi'\})=
\prod_{l=1}^2 \varrho^{(0)}_{stc}(q_l,t;\{\xi_l\}|q_l',t';\{\xi_l'\}),
\end{eqnarray}
where  taking into account (\ref{1.14}) we can write:
\begin{equation}
\label{4.02}
 \varrho^{(0)}_{stc}(q_l,t;\{\xi_l\}|q_l',t';\{\xi_l'\})=g_l^-\,
\exp\bigl\{\bar{A}_l(\textbf{u}^l(t)|q_l,q_l')\bigr\},
\end{equation}
 in addition:
$$
\bar{A}_l(\textbf{u}^l(t)|q_l,q_l')=-\int_{t_0}^{\,t}u_1(t')dt'+
A_l\bigl(\textbf{u}^{l}|q_l,q_l'\bigr),\quad
A_l= \frac{iu_1(t)}{2}\bigl(q_l^2-{q_l'}^2\bigr)
-\frac{u_2(t)}{2}\bigl(q_l^2+{q_l'}^2\bigr).
$$
To calculate the von Neumann entropy in the first step, we need to calculate the
reduced density matrix $\varrho^{(0,0)}(\textbf{q},\textbf{q}\,',t)$ (see definition 3).
Using the probability distribution $P(\textbf{u},t\vert\textbf{u}_0,t_0)$ we can construct
a continuous measure of the space $\mathbb{R}_{\{\bm\xi\}}$ and by the generalized
Feynman-Kac theorem calculate the functional integral (see in detail  \cite{AGev}):
\begin{eqnarray}
\label{4.03}
\varrho^{(0,0)}(\textbf{q},\textbf{q}{\,'},t)=\prod_{l=1}^2g_l^-
\int_{-\infty}^{+\infty}\int_{0}^{+\infty}
Q_{l}^{(0)}(u_1,u_2,t)\exp\bigl\{A_l\bigl(u_1,u_2|q_l,q_l'\bigr)\bigr\} du_1du_2,
\end{eqnarray}
where $Q_{l}^{(0)}(u_1,u_2,t)$ denotes the distribution function
of environmental fields, more precisely, SE fields formed under the influence
of QS in  the \emph{ground state}.

Note that the function $Q_{l}^{(0)}(u_1,u_2,t)$ satisfies the following
second-order \emph{partial differential equation} (PDE):
\begin{equation}
\label{4.04t}
\partial_t Q_{l}^{(n)}=\bigl\{\hat{L}_{l}\bigl(\Omega_{0l}(t), \epsilon_l|u_1,u_2,t\bigr)
-u_1(n+1)\bigr\}Q_{l}^{(n)},
\end{equation}
for  $n=0$.

Recall that $Q_{l}^{(0)}(u_1,u_2,t)$ describes the distribution function of
the fields of a environment or TB formed under the influence of a QS, which
is in an excited state characterized by a quantum number $n$. To solve the
equation (\ref{4.04t}), it is natural to use initial and boundary conditions
of the type:
\begin{eqnarray}
\label{4.05}
Q_{l}^{(0)}(u_1,u_2,t_0)=\prod_{l=1}^2\delta(u_l-u_{0l}),
\qquad
 Q_{l}^{(0)}(u_1,u_2,t)
\bigl|_{||\textbf{u}||\rightarrow\infty}\rightarrow 0.\qquad
\end{eqnarray}
Since  the function $Q_{l}^{(0)}(u_1,u_2,t)$ has a meaning of the probability density, it
can be normalized to unity:
$$
\bar{Q}_{l}^{(0)}(u_1,u_2,t)=\bigl(c_l^{(0)}(t)\bigr)^{-1}Q_{l}^{(0)}(u_1,u_2,t), \qquad
c_l^{(0)}(t)=\int_{-\infty}^{+\infty}\int_0^{+\infty} Q_{l}^{(0)}(u_1,u_2,t)du_1d u_2.
$$

Finally, substituting   (\ref{4.03}) for RDM in (\ref{2.07}), we obtain the
 von Neumann entropy:
\begin{equation}
\label{4.04a}
\Lambda_N(t) = -N^{(0)}_1(t)\Lambda_N^{(2)}(t)-N^{(0)}_2(t)\Lambda_N^{(1)}(t),
\end{equation}
where
$$
N^{(0)}_l(t)=(\Omega_l^- )^{1/2}\int_{-\infty}^{+\infty}\int_{0}^{+\infty} \frac{du_1du_2}{\sqrt{u_2}}
{Q}_{l}^{(0)}(u_1,u_2,t),
$$
$$
\Lambda_N^{(l)}(t)=\int_{-\infty}^{+\infty}\varrho^{(0)}(q_l,t)
\ln\bigl({\varrho^{(0)}(q_l,t)}\bigr)dq_l.
$$
As for the function $\varrho^{(0)}(q_l,t)$, then it has the form:
 $$
\varrho^{(0)}(q_l,t)=\int_{-\infty}^{+\infty}\int_{0}^{+\infty} du_1du_2
{Q}_{l}^{(0)}(u_1,u_2,t)\exp\bigl\{- {u_2q_l^2}\bigr\}.
$$
Finally, it is important to note that integration over the coordinates $q_1$ and $q_2$
in the obtained expressions cannot be performed analytically, which complicates the study
of the properties of the entropy function.

\subsection{Random entropy method}

Now we calculate the expression of entropy according to the definition (\ref{2.08}).

Substituting (\ref{4.01}) into expression (\ref{2.08}) and performing
simple calculations, we get the following expression (see \cite{Gev}):
\begin{equation}
\label{4.06}
\Lambda^{(0,0)}_G(t)=-N^{(0)}_1(t)\Lambda^{(0)}_2(t)-N^{(0)}_2(t)\Lambda^{(0)}_1(t),
\end{equation}
where
$$
\Lambda^{(0)}_l(t)=(\Omega_l^-)^{1/2}\int_{-\infty}^{+\infty}
\int_{0}^{+\infty}\frac{du_1 du_2}{\sqrt{u_2}} D_{l}(u_1,u_2,t).
$$
As for the function $D_{l}(\lambda_l,u_1,u_2,t)$, then this is a solution of the equation:
\begin{eqnarray}
\label{4.07}
\partial_t D_{l} =\hat{L}_{l}\,D_{l}-u_1 Q^{(0)}_{l}.
\end{eqnarray}
Note that to solve the equation (\ref{4.07}), the initial and boundary conditions of the
type (\ref{4.05}) are used. It is obvious that in the limit of thermodynamic equilibrium
the entropy should tend to the stationary limit:
\begin{equation}
\label{4.07a}
\Lambda^{(0,0)}_G({\bm{\bar\epsilon}}_l)=\lim_{t\to\,+\,\infty}\Lambda^{(0,0)}_G(t),
\qquad {\bm{\bar\epsilon}}_l=({\bar\epsilon}_1,{\bar\epsilon}_2).
\end{equation}

If we assume that in the limit $t\to+\infty$ the interaction between two coupled $1D$
oscillators vanishes or, equivalently, $\omega\to 0$, then ${\bar\epsilon}_1={\bar\epsilon}_2$.
However, this does not mean that the parameters $N^{(0)}_1$ with $N^{(0)}_2$ and $\Lambda_1$
with $\Lambda_2$ will be coincide. Moreover, as follows from the expression (\ref{4.03}),
if these oscillators have ever interacted, then the specified parameters will obviously
not be the same. Note that in both definitions of entropy (\ref{4.04a}) and (\ref{4.06}),
the entanglement of the states of two separate oscillators is obvious. In addition, it
is necessary to note that when the parameter
${\bm{\bar\epsilon}}_l=({\bar\epsilon}_1,{\bar\epsilon}_2)\to0$,
then the functions $ N^{(0)}_1 (t) $ and $ N^{(0)}_2 (t) $ tend to unity and, accordingly,
the entropy of separate oscillators and QS in whole should tend to zero, which will be
meet the condition of switching off an environment.

\section{E\lowercase{nergy levels and their occupancy after relaxation in} TB}
The energy spectrum is an important characteristic of a quantum system. Below we
will study the energy levels of the $1D$ oscillator after switching on the TB
and establishing thermodynamic equilibrium in the QS.
For the example, we will calculate the first several energy levels. Taking into
account (\ref{2.05}), it is  easy to obtain the expressions for the  mathematical
expectations of the energy levels. In particular, for the energy level of the
\emph{ground state} we get:
\begin{equation}
\mathcal{E}_{0}(\lambda_l,\mu)=\frac{1}{2}\bigl[1+K_0(\lambda_l,\mu)\bigr]\Omega_l^+,
\label{7.00}
 \end{equation}
where
$$
K_0(\lambda_l,\mu)=  \int_0^{+\infty} \int_{-\infty}^{+\infty}\frac{1}{\sqrt{\bar{u}_2}}
\biggl(-1+\frac{1+\bar{u}^2_1+\bar{u}^2_2}{2\bar{u}_2
d_l}\biggr)\widetilde{Q}^{(0)}_l (\lambda_l,\mu;\bar{u}_1,\bar{u}_2)d\bar{u}_1d\bar{u}_2.
$$
Recall that $d_l=\sqrt{\Omega^+_l/\Omega^-_l}$, in addition, the
 distribution function $\widetilde{Q}^{(0)}_l  $ is a solution
of the stationary dimensionless equation (\ref{4.04t}), which is formed in
the limit $\bar{t}\to +\infty $. Obviously, in the limit of $\lambda_l\to 0 $ and
$\mu=0$ for the energy level of the \emph{ground state} we should get the result:
$$\lim_{\lambda_l\to\,0}\mathcal{E}_{0}(\lambda_l,0)=(1/2)\Omega_l^+.$$
The latter, in turn, means that the distribution function $\widetilde{Q}^{(0)}_l $
 in this limit should have the following form:
\begin{equation}
\lim_{\lambda_l\to\,0}\,\widetilde{Q}^{(0)}_l (\lambda_l,0;\bar{u}_1,\bar{u}_2)
=\delta (\bar{u}_1)\delta(\bar{u}_2-\bar{u}_{02}),
\label{7.01}
 \end{equation}
where $\bar{u}_{02}=d_l\,\pm\sqrt{d^{\,2}_l-1}$.

In the case when $d_l>1$, obviously there are two solutions $\widetilde{Q}^{(0-)}_l$ and
$\widetilde{Q}^{(0+)}_l$ and, therefore, two energy levels  characterizing the
\emph{ground state}, which we will denote by $\mathcal{E}_{0}^-(\lambda_l,\mu)$
and $\mathcal{E}_{0}^+(\lambda_l,\mu)$, respectively.

Similarly, we can calculate the mathematical expectation of the energy level of
the first excited state:
\begin{equation}
\mathcal{E}_{1}(\lambda_l,\mu)=\frac{3}{2}\bigl[1+K_1(\lambda_l,\mu)\bigr]\Omega_l^+,
\label{7.02}
 \end{equation}
where
$$
K_1(\lambda_l,\mu)=  \int_0^{+\infty} \int_{-\infty}^{+\infty}\frac{1}{\bar{u}_2\sqrt{\bar{u}_2}}
\biggl(-1+\frac{1+\bar{u}^2_1+\bar{u}^2_2}{2\bar{u}_2
d_l}\biggr)\widetilde{Q}^{(2)}_l (\lambda_l,\mu;\bar{u}_1,\bar{u}_2)d\bar{u}_1d\bar{u}_2.
$$
 It is easy to verify that when  $d_l>1$, in this case also the energy level
is split into two sublevels  $\mathcal{E}_{1}^-(\lambda_l,\mu)$  and
$\mathcal{E}_{1}^+(\lambda_l,\mu)$, respectively.  Note that even in the case
of $d_l=1$, when  all sublevels disappear or, more precisely merge, the spectrum
of a quantum oscillator (QO) in the case under consideration is radically different
from the spectrum of QO without a environment. In particular, the equidistance
between energy levels is violated in this case.

Finally, we can calculate the population of different quantum levels as a function of temperature.
In particular, using the expression:
$$
M_n(\lambda_l,\mu)=Tr_{q_l}\bigl[Tr_{\xi_l}\bigl\{\varrho^{(n)}(q_l,t;\{\xi_l\}|q_l',t';\{\xi_l'\})\bigr\}\bigr],
$$
we can calculate the population of the first two energy levels. In particular, for the ground state,
the population level is determined by the expression:
\begin{equation}
M_0(\lambda_l,\mu)=  \int_0^{+\infty} \int_{-\infty}^{+\infty}\frac{1}{\sqrt{\bar{u}_2}}
Q^{(0)}_l (\lambda_l,\mu;\bar{u}_1,\bar{u}_2)d\bar{u}_1d\bar{u}_2,
\label{7.01ad}
\end{equation}
while for the first excited state it has the form:
\begin{equation}
M_1(\lambda_l,\mu)=  \int_0^{+\infty} \int_{-\infty}^{+\infty}\frac{1}{\bar{u}_2\sqrt{\bar{u}_2}}
Q^{(2)}_l (\lambda_l,\mu;\bar{u}_1,\bar{u}_2)d\bar{u}_1d\bar{u}_2.
\label{7.02ad}
\end{equation}

\section{E\lowercase{ntangled state caused by random influence of the thermostat}}
Performing coordinate transformations (\ref{1.03}), the original problem of
coupled oscillators is reduced to the problem of two noninteracting oscillators
in a random environment. Let the numbers 1 and 2, as  indicated above, denote
noninteracting oscillators
 in a random environment which wave states in the Hilbert spaces
$\mathbb{H}_1(\mathbb{R}^1\otimes \mathbb{R}_{\xi_1})$
and $\mathbb{H}_2(\mathbb{R}^1\otimes \mathbb{R}_{\xi_2})$
are denoted by the functions
$\Psi^{(1)}_{stc}\in L_2(\mathbb{R}^1\otimes \mathbb{R}_{\xi_1}) $
and $\Psi^{(2)}_{stc}\in L_2(\mathbb{R}^1\otimes \mathbb{R}_{\xi_2})$,
respectively.

 The Hilbert space of the composite system is the tensor product;
 $\mathbb{H}_\otimes=\mathbb{H}_1\otimes\mathbb{H}_2,$
while the state of the composite system is defined as:
\begin{equation}
 \Psi^{(1)}_{stc}\otimes\Psi^{(2)}_{stc}=\Bigl(\sum_{n}c_n^1|n\rangle_1\Bigr)
\otimes\Bigl(\sum_{m}c_m^2|m\rangle_2\Bigr)\,\in\mathbb{H}_\otimes .
 \label{5.01}
\end{equation}
In (\ref{5.01}) the vectors $|n\rangle_1=Y^{(1)}_{stc}(n|q_1,t;\{\xi_1\})$  and
$|m\rangle_2=Y^{(2)}_{stc}(m|q_2,t;\{\xi_2\})$ denote the
exact states of $1D$ quantum oscillators in the random environment
(see expression (\ref{1.14})), in addition, $c_n^1$ and $c_m^2$ are some complex
numbers with absolute values; $ |c_n^1|,\,|c_m^2| \leq 1$. Obviously, each set
of functions $\{|n\rangle_1\}$ and $\{|m\rangle_2\} $ forms an orthonormal
basis in the Hilbert spaces $\mathbb{H}_1$ and $\mathbb{H}_2$, respectively.

If the numbers $c_n^1$ and $c_m^2$ are not equal to zero, then in general
the separable states can be represented  as a direct product:
\begin{equation}
\Psi_{JS}=\sum_{n,m}c_{nm}|n\rangle_{1}\otimes |m\rangle_{2}, \qquad
c_{nm}=n^1_nc_m^2.
 \label{5.02}
\end{equation}
 where  $\Psi_{JS}$ denotes the wave function of JS.

Obviously,  based on the properties  (\ref{1.01})-(\ref{1.07}), it is easy
to show that in the extended space $\Xi\cong \mathbb{R}^2\otimes \mathbb{R}_{\{\bm\xi\}}$
the wave state $\Psi_{JS}$ is a separable. However, a reasonable question arises: what
happens if the wave function $\Psi_{JS}$ is averaged over the functional space
$R_{\{\bm\xi\}}$?  Obviously, from a physical point of view, this would mean
calculating the mathematical expectation  of the wave function of coupled
quantum oscillators taking into account the influence of a random environment:
\begin{equation}
\bar{\Psi}_{QS}({\bf{q}},t|n,m)=\mathbb{E}[\Psi_{JS}]=
\sum_{n,m}c_{nm}\overline{|n\rangle}_{1}\otimes \overline{|m\rangle}_{2},
\qquad n,m=0,1,2...,
 \label{5.03}
\end{equation}
where $\bar{\Psi}_{QS}=\mathbb{E}[\Psi_{JS}]=\langle\Psi_{JS}\rangle_{\mathbb{R}_{\{\bm\xi\}}}$
denotes the functional integration over the space  $\mathbb{R}_{\{\bm\xi\}}$.

For definiteness, we consider the case when $c_ {nm} = 0$ for   $n+ m\geq 1$, i.e.
consider  the case when both oscillators are in \emph{ground states}.
In this case, from (\ref{5.03}) for the wave function of quantum subsystem we
get the following expression:
\begin{equation}
\bar{\Psi}_{QS}({\bf{q}},t|0,0)=
 \overline{|0\rangle}_{1}\otimes \overline{|0\rangle}_{2},
 \label{5.03a}
\end{equation}
where
\begin{equation}
\overline{|0\rangle}_{l}=(g_l^-)^{1/2}\int_{-\infty}^{+\infty}
\int_{0}^{+\infty}\,\bar{\Upsilon}^{(\frac{1}{2},\frac{1}{2})}_{l}(u_1,u_2,t)
\exp\Bigl\{\frac{1}{2}\bigl(iu_1-u_2\bigr)q_l^2\Bigr\}du_1du_2,
 \label{5.05}
\end{equation}
describes the $ l$-\emph{th}  oscillator \emph{ground state} entangled with the environment.
As for the complex function $\bar{\Upsilon}^{(\frac{1}{2},\frac{1}{2})}_{l}(u_1,u_2,t)=
C_l(\frac{1}{2},\frac{1}{2})\Upsilon^{(\frac{1}{2},\frac{1}{2})}_{l}(u_1,u_2,t)$,
where $C_l(\frac{1}{2},\frac{1}{2})=const$,  it is the solution of the following PDE:
\begin{equation}
\label{5.07}
\partial_t\Upsilon^{(p,k)}_{l} =\bigl\{\hat{L}_{l}-(pu_1+iku_2)\bigr\} \Upsilon^{(p,k)}_{l},
\end{equation}
for the case   $(p,k)=1/2$.

 Obviously, the complex function $\Upsilon^{(p,k)}_{l}(u_1,u_2,t)$ cannot have a sense of
probability density. Rather, it can be interpreted as a wave function of the forming
\emph{small environment} (SE) that is closely related or, more accurately, entangled
with QS. In other words, the \emph{small environment} under the influence of QS \emph{is
quantized} and therefore the value $|\Upsilon^{(p,k)}_l(u_1,u_2,t)|^2$ should be
interpreted as a probability distribution of SE corresponding to the certain quantum
state of the QS. It is easy to show that for any values $(p, k)\in [0,\infty)$ the integral:
 $$
 1/C_l(p,k)=\int_{-\infty}^{+\infty} \int^{+\infty}_0
 |\Upsilon^{(p,k)}_l(u_1,u_2,t)|^2 du_1du_2 <\infty,
 $$
and therefore the function $\bar{\Upsilon}^{(p,k)}_{l}(u_1,u_2,t)$  can be
normalized to unity.

The equation (\ref{5.07}) can be represented as a system of two real equations:
\begin{eqnarray}
 \Biggl\{
\begin{array}{ll}
\label{5.08}
\partial_t R^{(p,k)}_{l} =\hat{L}_{l}R^{(p,k)}_{l}-\bigl(pu_1R^{(p,k)}_{l}-
ku_2 I^{(p,k)}_{l}\bigr),
\nonumber\\
\partial_t\, I^{(p,k)}_{l} =\,\hat{L}_{l}\,I^{(p,k)}_{l}-\bigl(pu_1I^{(p,k)}_{l}\,
+ku_2R^{(p,k)}_{l}\bigr),
\end{array}
\end{eqnarray}
where  $\Upsilon_l^{(p,k)}(u_1,u_2,t)=R^{(p,k)}_l(u_1,u_2,t) +iI_l^{(p,k)}(u_1,u_2,t)$.

It is easy to see that the system of equations (\ref{5.08}) is symmetric with respect
to permutations $R^{(p,k)}_{l}\to I^{(p,k)}_{l}$ and $I^{(p,k)}_{l}\to - R^{(p,k)}_{l}$.
On the other hand, this means that these solutions pass to each other as a result of
coordinate transformations. In particular, we can establish the following
general properties to which these solutions should satisfy:
\begin{eqnarray}
R^{(p,k)}_{l}(u_1,u_2,t)=\,I^{(p,k)}_{l}(-u_1,u_2,t)=-I^{(p,k)}_{l}( u_1,u_2,t),
\nonumber\\
I^{(p,k)}_{l}(u_1,u_2,t)=R^{(p,k)}_{l}(-u_1,u_2,t)=-R^{(p,k)}_{l}(u_1,u_2,t).
 \label{5.08a}
\end{eqnarray}
Given the properties (\ref{5.08a}), we can separate the equations by writing them in a
 mutually independent form:
\begin{eqnarray}
\label{5.08}
\partial_t R^{(p,k)}_{l} =\hat{L}_{l}R^{(p,k)}_{l}- \bigl(pu_1+ku_2\bigr)R^{(p,k)}_{l},
\nonumber\\
\partial_t\, I^{(p,k)}_{l} =\hat{L}_{l}\,I^{(p,k)}_{l}-\bigl(pu_1 -ku_2 \bigr)\,I^{(p,k)}_{l}.\,
\end{eqnarray}
To solve each of these partial differential equations, one can use initial and boundary
 conditions of the type (\ref{3.06}) or (\ref{4.05}).

For quantum communications, entangled states consisting of various vector states
are of particular interest. In particular, we can construct a quantum gate with
the following four \emph{Bell states}:
\begin{eqnarray}
\label{5.08k}
\Psi_{JS}^{\mp}=\frac {1}{\sqrt{2}}\bigl\{{|0\rangle}_{1}\otimes{|0\rangle}_{2}
\mp{|1\rangle}_{1}\otimes{|1\rangle}_{2}\bigr\},
 \nonumber\\
\Phi_{JS}^{\mp}=\frac {1}{\sqrt{2}}\bigl\{{|0\rangle}_{1}\otimes{|1\rangle}_{2}
\mp{|1\rangle}_{1}\otimes{|0\rangle}_{2}\bigr\}.
\end{eqnarray}

Conducting functional integration over these clear states, we obtain the following
mathematical expectations for Bell entangled states:
\begin{eqnarray}
\bar{\Psi}^{\mp}_{QS}(q_1,q_2,t)=\mathbb{E}\bigl[\Psi_{JS}^{\mp}\bigr]=\frac{1}{\sqrt{2}}
\bigl\{\overline{|0\rangle}_{1}\otimes\overline{|0\rangle}_{2}
\mp\overline{|1\rangle}_{1}\otimes\overline{|1\rangle}_{2}\bigr\},
\nonumber\\
\bar{\Phi}^{\mp}_{QS}(q_1,q_2,t)=\mathbb{E}\bigl[\Phi_{JS}^{\mp}\bigr]=\frac {1}{\sqrt{2}}
\bigl\{\overline{|0\rangle}_{1}\otimes\overline{|1\rangle}_{2}
\mp\overline{|1\rangle}_{1}\otimes\overline{|0\rangle}_{2}\bigr\},
 \label{5.0z4}
\end{eqnarray}
where $\bar{\Psi}^{\mp}_{QS}(q_1,q_2,t)=\langle\Psi_{JS}^{\mp}\rangle_{\mathbb{R}_{\{\bm\xi\}}}$
and   $\bar{\Phi}^{\mp}_{QS}(q_1,q_2,t)=\langle\Phi_{JS}^{\mp}\rangle_{\mathbb{R}_{\{\bm\xi\}}}$,
in addition:
\begin{equation}
\label{5.0t4}
\overline{|1\rangle}_{l}=2(g^-_l)^{1/2}q_l\int_{-\infty}^{+\infty}\int_{0}^{+\infty}\,
\bar{\Upsilon}^{(\frac{3}{2},\frac{3}{2})}_{l}(u_1,u_2,t)
\exp\Bigl\{\frac{1}{2}\bigl(iu_1-u_2\bigr)\,q_l^2\Bigr\}du_1du_2.
\end{equation}
Recall that the wave function $\bar{\Upsilon}^{(\frac{3}{2},\frac{3}{2})}_{l}(u_1,u_2,t)$
is a solution of the equation (\ref{5.07})  for the case $(p,k)=3/2$, which is normalized to unity.

Note that the states (\ref{5.0t4}) differ from ordinary Bell states in that their constructions
includes nonorthogonal  basis functions of the corresponding Hilbert spaces as a result of
additional integration over TB:
$$
\overline{|0\rangle}_{1},\,\overline{|1\rangle}_{1} \in \bar{ \mathbb{H}}_1^{(1)}(\mathbb{R}^1)
=\bigl\langle \mathbb{H}_1(\mathbb{R}^1\otimes \mathbb{R}_{\xi_1})\bigr\rangle_{\mathbb{R}_{\xi_1}},
$$
and, correspondingly,
$$
\overline{|0\rangle}_{2},\,\overline{|1\rangle}_{2} \in  \bar{\mathbb{H}}_1^{(2)}(\mathbb{R}^1)
=\bigl\langle \mathbb{H}_2(\mathbb{R}^1\otimes \mathbb{R}_{\xi_2})\bigr\rangle_{\mathbb{R}_{\xi_2}}.
$$
Note that an important feature of the developed representation is the presence of a number
of parameters that allow organizing effective external control over the QS.


\section{T\lowercase{ransitions probabilities between different asymptotic quantum states}}
Let us consider the evolution of the QS under the influence of a random environment, taking into account
possible quantum transitions. For definiteness, we assume that the fluctuations of the environment continue
 for a finite time. We assume that in the time interval
 $t\in (-\infty,t_0]$, the random force $f_l(t)$ acts on QS, and on the time interval $t\in[t_0,+\infty)$
 this influence disappears, i.e.  $f_l(t)\equiv0$. It follows from the above that in the $t\to+\infty$
 limit the wave function $\Psi_{out}(\textbf{m}|\textbf{q},t)$ has the form:
 \begin{equation}
 \Psi_{stc}(\textbf{n}|\textbf{q},t;\{\bm\xi\})=\sum_{\bf{m}}\mathrm{C}_{{\bf{n}}\bf{m}}(t;\{\bm\xi\})
 \Psi_{out}(\textbf{m}|\textbf{q},t),
\label{6.01a}
\end{equation}
where $\Psi_{out}(\textbf{m}|\textbf{q}, t) $ is the stationary wave function QS in the asymptotic state $(out)$,
 which describes the quantum state of coupled oscillators for times $ t> t_0 $, which can be represented as follows:
\begin{equation}
\Psi_{out}(\textbf{m}|\textbf{q},t)
=\prod_{l=1}^{2}e^{-i(m_l+1/2)\Omega_{l}^{+}t}\phi_{m_l}(q_l),
\label{6.02}
\end{equation}
where
\begin{equation}
\phi_{m_l}(q_l)=\biggl(\frac{g_l^+}{2^{m_l}{m_l}!}\biggr)^{1/2}
e^{-(\Omega_{l}^{+}q_l^2)/2} \mathrm{H}_{m_l} \Bigl(\sqrt{\Omega_{l}^{+}}\,q_l\Bigr), \qquad
g_l^+=({\Omega_{l}^{+}}/{\pi})^{1/2}.
\label{6.03}
\end{equation}

\textbf{Definition 7.}  \emph{ The mathematical expectation of the  transition  probability
between $(in)$ and $(out)$ asymptotic states  will be determined as:}
\begin{equation}
W_{\textbf{n}\to \bf{m}}=\lim_{t\to+\infty}\mathbb{E}\bigl[\bigl|\mathrm{S}_{\textbf{nm}}
(\textbf{q},t;\{\bm\xi\})\bigr|^2\bigr]=\lim_{t\to\,+\infty}\bigl|Tr_{\{\bm \xi\}}Tr_{\textbf{q}}
\bigl[\mathrm{S}_{\textbf{nm}}(\textbf{q},t;\{\bm\xi\})\bigr]\bigr|^2,
 \label{6.04}
\end{equation}
\emph{where $\mathrm{S}_{\textbf{nm}}(\textbf{q},t;\{\bm\xi\})=\Psi_{stc}(\textbf{n}|\textbf{q},t; \{\bm\xi\})
\Psi_{out}^\ast (\textbf{m}|\textbf{q},t)$ denotes a random $\bf{S}$-matrix element.}

To perform analytical calculations, we can use the generating functions method, but
for random processes (see \cite{Baz}):
\begin{equation}
\Psi_{stc}(\bm{\alpha}|\textbf{q},t;
\{\bm\xi\})=\prod_{l=1}^2\sum_{n_l=0}^{\infty}\frac{\alpha^{\,n_l}}{\sqrt{n_l!}}
Y_{stc}(n_l|q_l,t;\{\xi_l\}),
 \label{6.05}
\end{equation}
and
\begin{equation}
\Psi_{out}(\bm{\beta}|\textbf{q},t)=\prod_{l=1}^2\sum_{m_l=0}^{\infty}\frac{{\beta}^{\,m_l}}{\sqrt{m_l!}}\,
e^{-i(m_l+1/2)\,\Omega_{l}^{+}t}\phi_{m_l}(q_l),
 \label{6.06}
\end{equation}
where $\bm{\alpha}=(\alpha_1,\alpha_2)$ and $\bm{\beta}=(\beta_1,\beta_2)$ are auxiliary  complex  variables.

Calculating the sum in the representation (\ref{6.05}) leads to an expression that is the
 product of two Gaussian wave packets:
\begin{eqnarray}
\Psi_{stc}(\bm{\alpha}|\textbf{q},t;\{\bm\xi\})=\prod_{l=1}^2(g_l^-)^{1/2}
\exp\Bigl\{-\frac{1}{2}\bigl(a_lq_l^2-2b_lq_l+c_l\bigr)\Bigr\},
 \label{6.07}
\end{eqnarray}
where $a_l$, $b_l$ and $c_l$ are random variables, which are defined by the following expressions:
\begin{equation}
a_l=-i {\dot{\xi_l}}\xi_l^{-1},\qquad b_l=\sqrt{2\Omega^-_l}{\alpha_l}{\xi_l}^{-1},\qquad c_l
= {\xi^\ast_l}\xi_l^{-1}\alpha_l^2+\ln\xi_l.
 \label{6.08}
\end{equation}
 It is easy to see that  the random wave packet (\ref{6.07}) in the  $(in)$ asymptotic state
 passes a determined quantum state:
\begin{eqnarray}
\Psi_{stc}(\bm{\alpha}|\textbf{q},t;
\{\bm\xi\})\bigl|_{t\to-\,\infty}=\Psi_{in}(\bm{\alpha}|\textbf{q},t)= \qquad\qquad\qquad
\nonumber\\
\prod_{l=1}^2(g_l^-)^{1/2}\exp\biggl\{-\frac{1}{2}\Bigl(\Omega_l^-q_l^2-2\sqrt{2\Omega_l^-}
\alpha_lq_le^{-i\,\Omega_l^-t}+\alpha_l^2e^{-i\,2\Omega_l^-t}+i\,\Omega_l^-t\Bigr)\biggr\}.
 \label{6.09}
\end{eqnarray}
Obviously, the generating function $\Psi_{out}(\bm{\beta}|\textbf{q},t)$ can be easily
found from the expression (\ref{6.09}) by making the following substitutions
 $\Omega_l^-\to \Omega_l^+$ and $\alpha_l\to\beta_l$.

Now we consider the following integral:
\begin{eqnarray}
\mathrm{J}_{stc}(\bm\alpha,\bm\beta\,|\,t;\{\bm\xi\})=\int\Psi_{stc}(\bm{\alpha}|\textbf{q},t;
\{\bm\xi\})\Psi^\ast_{out}(\bm{\beta}\,|\textbf{q},t)d\textbf{q}.
 \label{6.10}
\end{eqnarray}
Performing simple calculations, we get:
\begin{eqnarray}
\mathrm{J}_{stc}(\bm\alpha,\bm\beta\,|\,t;\{\bm\xi\})=\prod_{l=1}^2
\frac{(\Omega_l^-\Omega_l^+)^{{1}/{4}}}{\sqrt{\bar{a}_l}}
\exp\biggl\{\frac{\bar{b}_l^2}{4\bar{a}_l}-\bar{c}_l\biggr\},
 \label{6.11}
\end{eqnarray}
where the following designations are made:
$$
\bar{a}_l=\frac{1}{2}\bigl(a_l+\Omega_l^+\bigr),\qquad \bar{b}_l=b_l+
\sqrt{2\Omega^+_l}\beta_l {e^{i\Omega_l^+t}},
\qquad
\bar{c}_l=\frac{1}{2}\bigl(c_l+\beta^2_le^{i2\Omega_l^+t}-i\,\Omega_l^+t\bigr).
$$
For further calculations, it is useful to represent the generating function (\ref{6.11}) as
the following decomposition:
$$
\mathrm{J}_{stc}(\bm\alpha,\bm\beta\,|\,t;\{\bm\xi\})=\prod_{l=1}^2
\sum_{n_l,m_l=0}^{\infty}\frac{\alpha^{n_l}_l}{\sqrt{n_l!}}
\mathrm{C}_{n_lm_l}^l (t;\{\xi_l\})\frac{\beta^{m_l}_l}{\sqrt{m_l!}},
\qquad \mathrm{C}_{{\bf{n}}\bf{m}}=\prod_{l=1}^2\mathrm{C}_{n_lm_l}^l,
$$
from  which follows, that:
\begin{eqnarray}
\prod_{l=1}^2\mathrm{C}_{n_lm_l}^l(t,\{\xi_l\})=\prod_{l=1}^2\frac{1}{\sqrt{n_l!m_l!}}
\frac{\partial^{n_l+m_l}}{\partial \alpha_l^{n_l}\partial\beta_l^{m_l}}
 \mathrm{J}_{stc}(\bm\alpha,\bm\beta\,|\,t;\{\bm\xi\})\Bigl|_{\alpha_l=\beta_l=0}.
 \label{6.12}
\end{eqnarray}
Using (\ref{6.11}) and (\ref{6.12}), we can write the transition probability in the following form:
\begin{equation}
W_{\textbf{n}\to\textbf{m}}=
\lim_{t\to+\infty}\biggl\{\prod_{l=1}^2\bigl|Tr_{\{\xi_l\}}
\bigl[\mathrm{C}^l_{n_lm_l}(t;\{\xi_l\})\bigr]\bigr|^2\biggr\}.
 \label{6.13}
\end{equation}
It is easy to verify that the transition probability (\ref{6.13}) formally is a product
of the transitions probabilities of two one-dimensional oscillators. In other words, we can
represent the equation (\ref{6.13}) in a factored form:
\begin{equation}
W_{\textbf{n}\to\textbf{m}}= \prod_{l=1}^2  w^{(l)}_{n_l\to m_l},\qquad W^{(l)}_{n_l\to m_l}=
\lim_{t\to+\infty} \bigl|Tr_{\{\xi_l\}}\bigl[\mathrm{C}^l_{n_lm_l}(t;\{\xi_l\})\bigr]\bigr|^2.
 \label{6.13c}
\end{equation}
Thus, by calculating the transition probabilities of $1D$ quantum oscillator, we can
construct the corresponding transitions of 2$D$ oscillator.

For definiteness, we calculate a series of transition probabilities between
different asymptotic states $(in)$ and $(out)$. Taking into account (\ref{6.08}), (\ref{6.11})
and (\ref{6.13}),  we can  construct  explicit form of the functional integral and
calculate it using the generalized Feynman-Kac theorem (see \cite{}). In particular, for
 the transition probability between the \emph{ground states} of the $(in)$ and $(out)$
asymptotic channels, we get the following integral representation:
\begin{equation}
W^{(l)}_{0\to\,0}=\kappa_l \Biggl|
\int_{-\infty}^{+\infty} \int_{0}^{+\infty}\frac{\bar{\Upsilon}_l^{(\frac{1}{2},\frac{1}{2})}
(\lambda_l,\mu;\bar{ u}_1,\bar{u}_2)}{(1+\bar{u}_2 - i\bar{u}_1)^{\frac{1}{2}}}
 d\bar{u}_1d\bar{u}_2 \Biggr|^2,
 \label{6.14}
\end{equation}
where   the following designations are made:
$$  \lambda_l=\bar{\epsilon}_l/(\Omega_l^+)^3,\qquad \kappa_l=2(\Omega_l^-/\Omega_l^+)^{1/2},
 \qquad\bar{ u}_1=u_1/\Omega_l^+,\qquad\bar{u}_2=u_2/\Omega_l^+.$$
As for the stationary wave function
$\bar{\Upsilon}_l^{(\frac{1}{2},\frac{1}{2})}(\lambda_l,\bar{u}_1,\bar{u}_2)$, then it
is the  solution of the equation (\ref{5.07}) in the limit of $\bar{t}\to+ \infty$,
which in dimensionless form is written as:
\begin{equation}
 \partial_{\,\bar{t}}\bar{\Upsilon}_l^{(p,k)}=\bigl\{ \bar{\hat{L}}_{l}- (p\,\bar{u}_1+
ik\,\bar{u}_2) \bigr\}\bar{\Upsilon}_l^{(p,k)}, \qquad \bar{t}=\Omega_l^+t,
 \label{6.a14}
\end{equation}
where
 $$
\bar{\hat{L}}_l= \lambda_l\Bigl(\frac{\partial^{\,2}}{\partial\bar{u}_1^2}\,+\mu
 \frac{\partial^{\,2}}{\partial \bar{u}_2^2}\,\Bigr)+ \frac{\partial
}{\partial\bar{u}_1}\bigl(\bar{u}_1^2-\bar{u}_2^2+\bar{\Omega}^2_{0l}(t)\bigr)+2\bar{u}_1
\frac{\partial}{\partial\bar{u}_2}\bar{u}_2,\qquad \bar{\Omega}_{0l}(t)
=\frac{\Omega_{0l}(t)}{\Omega_{l}^+}.
$$

Similarly, we can calculate probabilities of other transitions of $1D$ oscillator. In
particular the first few  transitions can be represented in the form:
\begin{eqnarray}
W^{(l)}_{1\to\,1}= \kappa^3_l \Biggl|
\int_{-\infty}^{+\infty} \int_{0}^{+\infty} \frac{\bar{\Upsilon}_l^{(\frac{3}{2},\frac{3}{2})}
(\lambda_l,\mu;\bar{ u}_1,\bar{u}_2)}
{(1+\bar{u}_2 - i\bar{u}_1)^{\frac{3}{2}}}
 d\bar{u}_1d\bar{u}_2 \Biggr|^2,\qquad\qquad\qquad\qquad\qquad\quad
\nonumber\\
W^{(l)}_{0\to\,2}=\kappa_l \Biggl|
\int_{-\infty}^{+\infty} \int_{0}^{+\infty} \frac{1-\bar{u}_2+i\bar{u}_1}
{(1+\bar{u}_2 - i\bar{u}_1)^{\frac{3}{2}}}
\bar{\Upsilon}_l^{(\frac{1}{2},\frac{1}{2})}(\lambda_l,\mu;\bar{ u}_1,\bar{u}_2)
d\bar{u}_1d\bar{u}_2 \Biggr|^2,\qquad\qquad \,\,
\nonumber\\
W^{(l)}_{2\to\,0}=\kappa_l \Biggl|
\int_{-\infty}^{+\infty} \int_{0}^{+\infty}\biggl[\kappa^2_l
\frac{\bar{\Upsilon}_l^{(2,2)}(\lambda_l,\mu;\bar{ u}_1,\bar{u}_2)}
{(1+\bar{u}_2 - i\bar{u}_1)^{\frac{3}{2}}}
-\frac{\bar{\Upsilon}_l^{(2,0)}(\lambda_l,\mu;\bar{ u}_1,\bar{u}_2)}{(1+\bar{u}_2 -
i\bar{u}_1)^{\frac{1}{2}}}\biggr] d\bar{u}_1d\bar{u}_2 \Biggr|^2,
 \label{6.14z}
\end{eqnarray}
where the wave function $\bar{\Upsilon}_l^{(p,k)}(\lambda_l,\mu;\bar{u}_1,\bar{u}_2)$
denotes the stationary solution of the equation (\ref{6.a14})
in the limit of $\bar{t}\to+\infty.$

In particular, as follows from the expressions of transition probabilities (\ref{6.14z}),
only transitions between states with the same parity are possible. However, the most
important and unexpected result in this case is that when the frequency is constant, i.e.
$\Omega_{0l}(t)=const$, the  detailed balance between the quantum levels
$W_{nm}\neq W_{mn}$ is  disturbed that is the cornerstone law of standard
quantum mechanics.  In particular, we can verify this by comparing the following two
transitions $W^{(l)}_{0\to\, 2} $ and $W^{(l)}_{2\to\, 0}$.

Now it is important to show that in the limit of turning off the environment, the
probabilities of transitions pass into regular well-known expressions. For the example,
let us rewrite the expression for the \emph{ground stat-ground stat} transition
(\ref{6.14}) in the form:
 \begin{eqnarray}
W^{(l)}_{0\to\, 0}= \kappa_l
\int_{-\infty}^{+\infty}
\int_{0}^{+\infty}
\bigl[1-\varrho(\bar{u}_1,\bar{u}_2)\bigr]^{1/4}\,\bar{\Upsilon}_l^{(\frac{1}{2},\frac{1}{2})}
(\lambda_l,\mu;\bar{u}_1,\bar{u}_2)\exp\bigl\{-i\varphi(\bar{u}_1,\bar{u}_2)\bigr\}
d\bar{u}_1d\bar{u}_2,
 \label{6.16}
\end{eqnarray}
where
$$
\varrho(\bar{u}_1,\bar{u}_2)=1-\frac{\bar{u}_1^2+\bar{u}_2^2+2\bar{u}_2}{\bar{u}_1^2+(1+\bar{u}_2)^2},\quad
\varphi(\bar{u}_1,\bar{u}_2)=\frac{1+\bar{u}_2}{\sqrt{\bar{u}_1^2+(1+\bar{u}_2)^2}},
\qquad 0\leq\varrho(\bar{u}_1,\bar{u}_2)\leq1.
$$
To pass to the well-known 1$D$  problem of a parametric oscillator, it is obviously necessary to put
$\mu=0$ and $\lambda_l\to 0$.  The latter, in turn, implies that the imaginary part of the solution
 $\Upsilon_l^{(\frac{1}{2},\frac{1}{2})}$ is zero and, consequently, it is
 necessary to replace $\Upsilon_l^{(\frac{1}{2},\frac{1}{2})} \to R_l^{(\frac{1}{2},\frac{1}{2})}$
and, in addition, the  stationary solution
$R_l^{(\frac{1}{2},\frac{1}{2})}(\lambda_l,\mu;\bar{u}_1,\bar{u}_2)$  must satisfy the condition:
\begin{equation}
\lim_{\lambda_l\to\, 0}R_l^{(\frac{1}{2},\frac{1}{2})}(\lambda_l,\mu;\bar{u}_1,\bar{u}_2)=\lambda_l^{-1}
\delta(\bar{u}_1)\delta(\bar{u}_2-\bar{u}_{02}), \qquad
\bar{u}_{02}=-1+\frac{1}{\sqrt{1-\rho}},
\label{6.1k6}
 \end{equation}
where $\rho= \bigl|{C_2^{(l)}}/{C_1^{(l)}}\bigr |^2$.

Recall that the coefficients
$C_1^{(l)}$ and $C_2^{(l)}$ are found from the solution  of the classical  equation
(\ref{1.13}) with the regular frequency $\Omega_{0l}(t)$, in the limit $t\to+\infty$,
when a classical oscillator transit in $(out)$ channel, and the solution in this case
is has the form; $\xi_{0l}(t)\sim\, C_{1}^{(l)}e^{i\Omega_{l}^+ t}-C_{2}^{(l)}e^{-i\Omega_{l}^+ t}$.

Finally, substituting (\ref{6.1k6}) in (\ref{6.16}), we get a well-known result for the transition
probability of $1D$ parametric oscillator; $W^{(l)}_{0\to\,0}=\sqrt{1-\rho}$.
Note that in the same way for $\lambda_l\to 0 $ we can pass to the known regular values
(see \cite{Baz})  for other transitions (\ref{6.14z}).


\section{C\lowercase{onclusion}}
The main purpose of this study was to develop an analytical model of the joint system
``quantum subsystem+environment (universe)", which would allow a self-consistent study
of the evolution of both the quantum subsystem and its environment. To implement this
idea, as a basic equation for describing  JS, we chose a Langevin-Schr\"{o}dinger
type SDE, for which the Schr\"{o}dinger equation plays the role of a local correspondence.
In other words, we suggest that both equations, SDE and the Schr\"{o}dinger equation match
at small time intervals.

In particular, for a deeper understanding of the problems of quantum foundations,
including the connection of \emph{quantum thermodynamics with the first principles of quantum
mechanics}, concrete, \emph{exactly constructed models} can be very informative and useful.
In this article, we focused on the problem of two coupled oscillators immersed
in a thermostat, or on the ``CQO + TB" problem, which is an ideal model of a bi-molecular
quantum reacting gas. We have shown that within the framework of this model, all the
statistical parameters of the quantum subsystem can be constructed in a closed form in
the form of two-dimensional integral representations and solutions of second-order PDEs.
However, in our opinion, it was very important to prove the formation of a \emph{quantized
small environment} (QSE) as a result of complex nonlinear self-organization processes in the JS.
The physical meaning of QSE can be interpreted as a continuation of the quantum subsystem or,
more precisely, its \emph{quantum halo}, which contains information about QS.

In the work, the time-dependent quantum entropy is calculated for CQO in the ground state.
 It was shown that the von Nueman quantum entropy (see expressions (\ref{2.07}) and (\ref{4.03}))
and the generalized quantum entropy taking into account the relaxation of an environment
(\ref{4.06}), are different. Both expressions of entropy coincide only when the
influence of the environment is considered a small perturbation.

Note that from the generalized entropy expression (\ref{4.06}) follows that the environment
makes the quantum subsystem inseparable. Even when the QS splits into two parts, and its parts
are removed on infinity, a non-potential interaction arises, characteristic of entangled
quantum states. The considered problem gives us the opportunity to study in detail and
deeply the role of the medium in the phenomenon of entanglement of spatially isolated
quantum subsystems, and also allows us to organize effective control of entanglement
properties through environmental parameters. We expect that a  simulation of expressions
(\ref{5.05})  can give important information on a role of entangling in the
process of thermal relaxation of an environment and degree of violation of the
\emph{basic principle of statistical physics}- on the \emph{equiprobability of statistical
states} \cite{Popescu}.

In Section VIII, the probabilities of transitions between different asymptotic states of
CQO are calculated explicitly taking into account the influence of the TB. The latter allows
one to construct kinetic equations and directly simulate the population of levels of the
QS, which is very important for testing the hypothesis of micro-canonical distribution in a
quantum ensemble under thermal equilibrium.

Finally, it should be noted that this study may also shed new light on some fundamental
problems of the quantum foundations and quantum statistical mechanics, such as the recent
debate on the possibility of violating certain thermodynamic laws, in particular the second
law of thermodynamics \cite{Opat,Cap}.


\end{document}